\documentclass[draft,grl]{AGUTeX}

  \usepackage[dvips,dvipdf]{graphicx}
  \setkeys{Gin}{draft=false}
\usepackage{rotating}
\usepackage{tabularx}
\usepackage{color} 
\usepackage{tablefootnote}

\newcommand{\KWmm}{K~(W~m$^{-2}$)$^{-1}$}
\newcommand{\be}{\begin{equation}}
\newcommand{\ee}{\end{equation}}
\newcommand{\bae}{\begin{eqnarray}}
\newcommand{\eae}{\end{eqnarray}}
\newcommand{\bse}{\begin{subeqnarray}}
\newcommand{\ese}{\end{subeqnarray}}

\def\deg{ ^{\circ}}

\authorrunninghead{VON DER HEYDT ET AL.}

\titlerunninghead{STATE-DEPENDENT CLIMATE SENSITIVITY}

\authoraddr{Corresponding author: A. S. von der Heydt,
Institute for Marine and Atmospheric research Utrecht, Utrecht University, P. O. Box 80.005, 3584TA Utrecht, The Netherlands.
(a.s.vonderheydt@uu.nl)}

\begin{document}

%% ------------------------------------------------------------------------ %%
%
%  TITLE
%
%% ------------------------------------------------------------------------ %%

\title{On the state dependency of fast feedback processes in (palaeo) climate sensitivity}
%
% e.g., \title{Terrestrial ring current:
% Origin, formation, and decay $\alpha\beta\Gamma\Delta$}
%
%% ------------------------------------------------------------------------ %%
%
%  AUTHORS AND AFFILIATIONS
%
%% ------------------------------------------------------------------------ %%

%Use \author{\altaffilmark{}} and \altaffiltext{}

% \altaffilmark will produce footnote;
% matching \altaffiltext will appear at bottom of page.

 \authors{A. S. von der Heydt,\altaffilmark{1}
 P. K\"ohler,\altaffilmark{2} R. S. W. van de Wal,\altaffilmark{1}
 and H. A. Dijkstra\altaffilmark{1}}

\altaffiltext{1}{Institute for Marine and Atmospheric research Utrecht, Utrecht University, P. O. Box 80.005, 3584TA Utrecht, The Netherlands.}

\altaffiltext{2}{Alfred-Wegener-Institut, Helmholtz-Zentrum f\"ur Polar- und Meeresforschung, P. O. Box 12 01 61, 27515 Bremerhaven, Germany}

%% ------------------------------------------------------------------------ %%
%
%  ABSTRACT
%
%% ------------------------------------------------------------------------ %%

% >> Do NOT include any \begin...\end commands within
% >> the body of the abstract.

\begin{abstract}
Palaeo data have been frequently used to determine the equilibrium (Charney) climate sensitivity $S^a$, and --- if slow feedback processes (e.g. land ice-albedo) are adequately taken into account --- they indicate a similar range as estimates based on instrumental data and climate model results. Most studies implicitly assume the (fast) feedback processes to be independent of the background climate state, e.g., equally strong during warm and cold periods. Here we assess the dependency of the fast feedback processes on the 
background climate state using data of the last 800~kyr and a conceptual climate model for interpretation. Applying a new method to account for background state dependency, we find $S^a~=~0.61\pm~0.06$~\KWmm\ using the latest LGM temperature reconstruction and significantly lower climate sensitivity during glacial climates. 
Due to uncertainties in reconstructing the LGM temperature anomaly, $S^a$ is estimated in the range $S^a$~=~0.55~--~0.95~\KWmm.
\end{abstract}

%% ------------------------------------------------------------------------ %%
%
%  BEGIN ARTICLE
%
%% ------------------------------------------------------------------------ %%

% The body of the article must start with a \begin{article} command
%
% \end{article} must follow the references section, before the figures
%  and tables.

\begin{article}

%% ------------------------------------------------------------------------ %%
%
%  TEXT
%
%% ------------------------------------------------------------------------ %%

\section{Introduction}
The  Charney climate sensitivity $S^a$ is determined by  fast  feedbacks, i.e. those with a response time scale faster than a typical forcing time scale (usually taken as $\sim$100 years for the anthropogenic $\rm{CO}_2$ increase \citep{Charney1979b,Knutti2008,Rohling_palaeosens2012}). Recently, a systematic approach has been proposed to determine $S^a$ from palaeoclimate data by correcting the values of the specific climate sensitivity $S_{[{\rm CO}_2]}$ caused by the radiative forcing of atmospheric  $\rm{CO}_2$ changes for the slow feedbacks such as land-ice albedo \citep{Rohling_palaeosens2012}. This approach has revealed values of $S^a$ within a range  of 
$0.6 - 1.3$~\KWmm\  at the 68\% probability level for the last 65 million years, which is similar to the range estimated from the CMIP5 climate model ensemble \citep{Vial2013}. 
There are, however, several assumptions made in order to determine the estimates of $S^a$ from palaeo records. One of them is that the strength of the  fast feedbacks is  independent of the background state of the climate system.  Many studies mention that this assumption may be unrealistic \citep{Senior2000, Crucifix2006,Andrews2008,Yoshimori2011}, but the effect of the background state dependency on the values of $S^a$ has not been quantified. 

%Specific climate sensitivities: 
Climate sensitivity is determined by $S=\frac{\Delta T}{\Delta R}$, where $\Delta T$ is the global mean temperature change and $\Delta R$ is the change in radiative forcing. 
The specific climate sensitivities $S_{[X,Y, ...]}$ \citep{Rohling_palaeosens2012} depend on whether a process is considered as forcing or feedback, where the subscripts $X,Y,...$ denote the forcings (SI, section S2), e.g.
%\begin{linenomath*}
\begin{eqnarray}
S_{[{\rm CO}_2]}&=\frac{\Delta T}{\Delta R_{[{\rm CO}_2]}}, \label{eq:sco2}\\
S_{[{\rm CO}_2,LI]}&=\frac{\Delta T}{\Delta R_{[{\rm CO}_2]}+\Delta R_{[LI]}}. 
\label{eq:sco2li}
\end{eqnarray}
%\end{linenomath*} 
Here, $\Delta R_{[{\rm CO}_2]}$ and $\Delta R_{[LI]}$ are, respectively, the radiative forcing contributions of $\rm{CO}_2$ and of surface albedo changes caused by land-ice (LI). The specific climate sensitivity $S_{[{\rm CO}_2]}$ derived from palaeo data is based on reconstructed values for $\Delta T$ and $\Delta R_{[{\rm CO}_2]}$. To estimate the Charney sensitivity $S^a$, the values of $S_{[{\rm CO}_2]}$ need to be corrected for the slow feedback processes or forcings other than CO$_2$ \citep{Rohling_palaeosens2012}. Therefore, reconstructions of land-ice area and other slow processes are necessary as well.  

In this paper, we revisit the concept of background state dependency and analyse palaeoclimate data from the glacial-interglacial transitions during the Late Pleistocene \citep{Koehler2010}.
$S^a$ was calculated previously from the Last Glacial Maximum (LGM) part of these data and corrected for state dependency based on a single climate model \citep{Hargreaves2007}. However, it has been argued that such corrections are highly model-dependent \citep{Crucifix2006}. Here we suggest a new method for estimating climate sensitivity from palaeo data in order to account for background state dependency of the fast feedbacks.

\section{A conceptual model to estimate climate sensitivity}
%Conceptual climate model

\begin{figure}
\noindent\includegraphics[width=\textwidth]{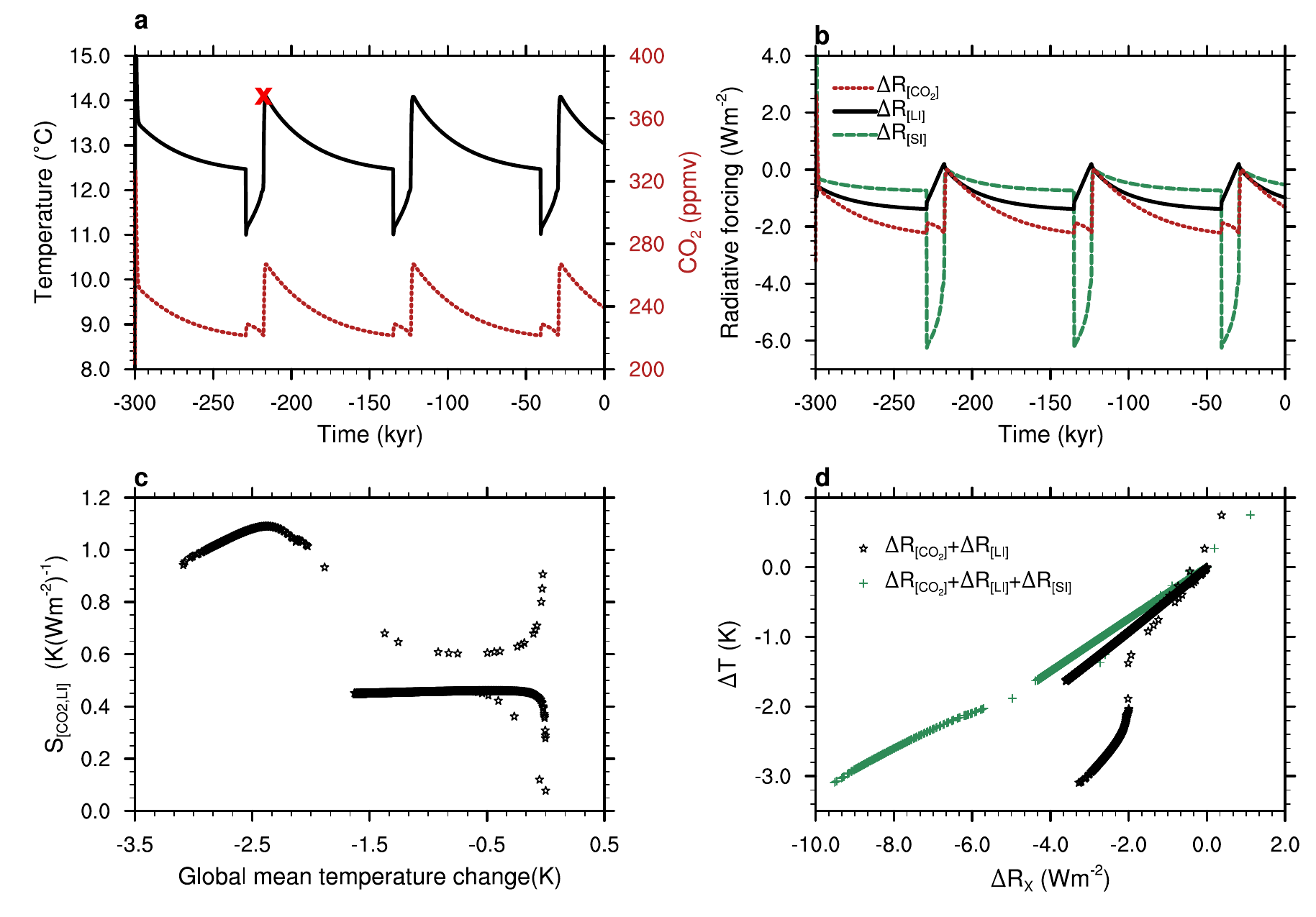}
\caption{Climate sensitivity  based on the conceptual climate model results. (a) The 300 kyr simulation shows three glacial-interglacial cycles with a peak-to-peak global mean temperature difference of 3.2~K and atmospheric $\rm{CO}_2$ variations of 50$~$ppmv; red cross marks the reference temperature for an interglacial;(b) Radiative perturbations due to atmospheric $\rm{CO}_2$ changes (red short dashed line), land-ice cover (black solid line) and sea-ice cover (green long dashed line); (c) Specific climate sensitivity $S_{[{\rm CO}_2,LI]}$ calculated in the traditional way with a fixed reference climate (red cross in (a));  (d) Temperature anomalies $\Delta T$ versus radiative forcings: The (local) slopes determine the state dependent specific climate sensitivities.}
\label{f:sensitivity_model}
\end{figure}

To understand the characteristics of the state dependency of the fast feedbacks we use 
a conceptual climate model  \citep{Gildor2001a}, which has been shown to simulate the 
glacial-interglacial transitions due to the so-called sea-ice switch mechanism 
(SI, section S1, for a detailed model description). This model contains
 two main feedbacks, the fast sea-ice albedo and the slow land-ice albedo 
feedbacks. 
%On fast to intermediate time scales there is one additional ÓfeedbackÓ in the radiative balance due to heat exchange between ocean and atmosphere.
To estimate the climate sensitivity, two climate states are compared with a temperature difference $\Delta T$ and radiative 
changes due to the different processes $\Delta R_{[X]}$ (with $X$ indicating the process) which can  
be explicitly computed in the model (Fig.~\ref{f:sensitivity_model}a,b). 
As the only slow feedback process in the model is the land ice-albedo feedback, the value of $S_{[{\rm CO}_2,LI]}$ already accurately approximates $S^a$. 

When determining climate sensitivity from palaeo time series the differences in temperature and radiative forcing are taken usually with respect to a fixed reference climate, mostly the preindustrial climate. Fig.~\ref{f:sensitivity_model}c shows the model climate sensitivity determined by differences with respect to an interglacial climate (red cross in Fig.~\ref{f:sensitivity_model}a) from a 300 kyr simulation. In this figure, we can clearly distinguish two temperature regimes, with higher climate sensitivity values during cold periods and lower values during warm periods. 
The sea ice-albedo feedback, which is the dominant fast feedback in the model explains the higher sensitivity during glacial times: there is more sea ice available to melt under a doubling of $\rm{CO}_2$, indicating a stronger feedback and a higher sensitivity.

However, by this approach, the climate sensitivity is linearised between two reference states. In case of a non-linear climate system the error made by this linearisation becomes the larger the more distant (in terms of temperature) climates are compared. Hence, we propose an alternative approach to determine the state dependency of climate sensitivity more accurately, using local slopes (SI, section S3). 
According to eq.~(1b), $S_{[{\rm CO}_2,LI]}$ (in the model approximating $S^a$) is the slope in a graph showing $T$ versus $(R_{[{\rm CO}_2]}+\Delta R_{[LI]})$ as shown in Fig.~\ref{f:sensitivity_model}d. 
Under the assumption of no state dependency (linear climate system) we expect a linear relation between $\Delta T$ and $(\Delta R_{[{\rm CO}_2]}+\Delta R_{[LI]})$, i.e. with a constant slope. The two regimes visible in Fig.~\ref{f:sensitivity_model}d (black symbols) with different (local) slopes are an expression of the fact that the fast feedbacks depend on the background climate state. $S^a$ is generally higher for cold climates than for warm climates, however, the values of $S^a$ for cold climates as determined from the local slope in Fig.~\ref{f:sensitivity_model}d vary from $S^a\simeq 0.5$\KWmm\ to almost infinity at the discontinuous point close to $\Delta R_{[CO2]}+\Delta R_{[LI]}\simeq -2$~Wm$^{-2}$. Such large variations do not appear in Fig.~\ref{f:sensitivity_model}c when $S^a$ is determined from differences with respect to a fixed reference climate.  

We can take the approach one step further by correcting the sensitivity also for the (fast) sea-ice albedo feedback, i.e., determining the specific sensitivity $S_{[{\rm CO}_2,LI,SI]}=\Delta T/(\Delta R_{[{\rm CO}_2]}+\Delta R_{[LI]}+\Delta R_{[SI]})$. In the model this should remove nearly all state dependency as there are no other feedbacks. The relation between $\Delta T$ and $(\Delta R_{[{\rm CO}_2]}+\Delta R_{[LI]}+\Delta R_{SI})$ (green symbols in Fig.~\ref{f:sensitivity_model}d) is still not exactly linear, indicating that also the remaining feedback depends on the background state, however in the opposite way than the sea ice-albedo feedback, with a slightly smaller sensitivity during glacial periods. This might be also an expression of the fact that on the time scales considered the ocean-atmosphere heat exchange (also represented in the conceptual model) is not exactly in equilibrium (SI, section S2). 

In conclusion, the model results show that climate sensitivity should be determined from local slopes of the temperature--radiative forcing relation to take into account background state dependency of the fast feedbacks. 

\section{Data of the last 800 kyr}

To estimate the dependency of the fast feedbacks on the background state from palaeo data, 
we use a compilation of several environmental records and model-based 
derived variables over the last 800 kyr \citep{Koehler2010,Rohling_palaeosens2012}. 

\begin{figure}
\noindent\includegraphics[width=\textwidth]{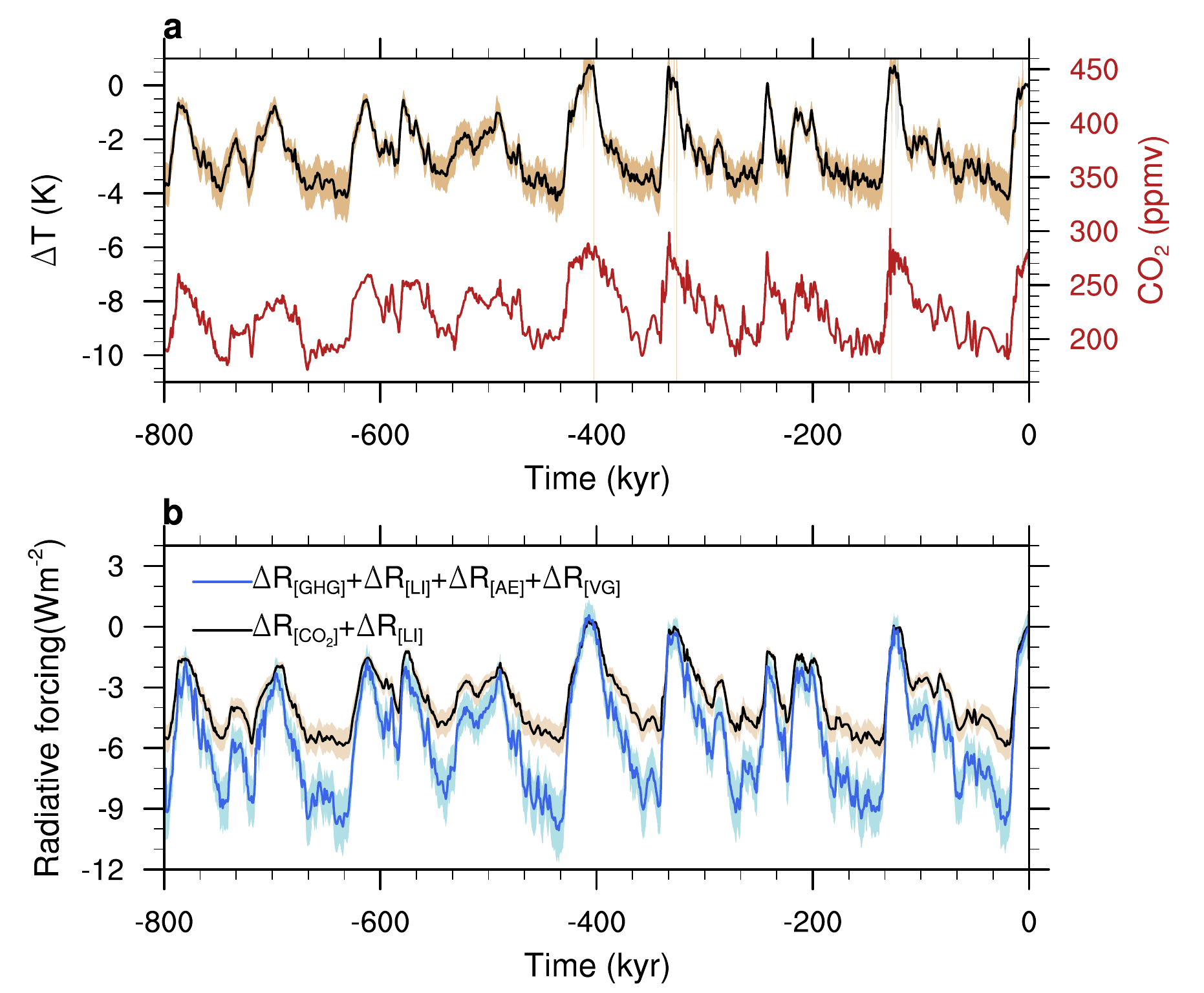}
\caption{Data over the last 800 kyr used to estimate climate sensitivity. Shaded areas around the curves indicate uncertainty intervals ($\pm1\sigma$);  (a) Global mean temperature anomalies with respect to $T^{pi} =$~286.5~K (black line) considering a global cooling at LGM of $\Delta T=-4.0$~K and $\rm{CO}_2$ (red line) records; (b) Radiative forcing due to atmospheric $\rm{CO}_2$ and land-ice cover (black line) and due to all known and reconstructed forcings, i.e. GHG, LI, AE and VG (blue line). }
\label{f:sensitivity_data1}
\end{figure}

Our estimated global temperature anomalies are based on (i) the deconvolution  of the benthic $\delta^{18}O$-stack \citep{Lisiecki2005,Bintanja2005} into a northern hemispheric land ($40^\circ-80^\circ$~N) temperature anomaly $\Delta T_{NH}$ combined with a constant polar amplification factor $\alpha_{NH}=3.75\pm 0.35$ ($\pm 1 \sigma$) and (ii) an Antarctic temperature anomaly $\Delta T_{ANT}$ from the EPICA Dome C data \citep{Jouzel2007} with polar amplification factor $\alpha_{ANT}=2.25\pm 0.25$, to match the most recent global mean temperature reconstruction at LGM of $-4.0$~K \citep{Annan2013} by $\Delta T = (\frac{\Delta T_{NH}}{\alpha_{NH}}+\frac{\Delta T_{ANT}}{\alpha_{ANT}})/2$ (Fig.~\ref{f:sensitivity_data1}a). The resulting polar amplification factors are high compared to climate model results \citep{MassonDelmotte2006}. 
In the SI, section S5, a similar analysis as below is shown but then based on the LGM temperature reconstruction by \citet{SchneidervonDeimling2006} and a polar amplification factor of $\alpha=2.75$ which is closer to model estimates of polar amplification \citep{MassonDelmotte2006}. The $\rm{CO}_2$ reconstruction \citep{Petit1999, Monnin2001, Siegenthaler2005, Luethi2008} (Fig.~\ref{f:sensitivity_data1}a) is used to calculate radiative forcing changes due to $\rm{CO}_2$ \citep{Myhre1998}. Radiative forcing changes due to albedo of land-ice coverage are calculated from the land-ice area reconstructions \citep{Bintanja2005} in line with \citet{Rohling_palaeosens2012}. 

Our data set contains uncertainties of all variables as estimated previously by \citep{Koehler2010,Rohling_palaeosens2012} and is interpolated to 100-year time steps. $\Delta T$ and $\Delta R$ are calculated as anomalies with respect to preindustrial values (see SI, section S4 for a detailed description of the uncertainty analysis). 

The temperature record together with the radiative forcing changes due to  $\rm{CO}_2$ and reconstructed land-ice albedo variations (Fig.~\ref{f:sensitivity_data1}) allows calculating the specific sensitivities  $S_{[{\rm CO}_2]}$ and $S_{[{\rm CO}_2,LI]}$.  In the traditional way with a fixed reference climate, i.e. the preindustrial climate, we find values $S_{[{\rm CO}_2]} = 2.05 \pm 0.75$~\KWmm \ and $S_{[{\rm CO}_2,LI]} = 0.70 \pm 0.18$~\KWmm\ from the complete data set. The latter is similar to results from a model ensemble for the LGM \citep{Hargreaves2012}. 
In the real climate system, the land-ice albedo feedback is not the only slow feedback, and therefore, $S_{[{\rm CO}_2,LI]}$ is only an approximation of $S^a$. Other factors such as dust, vegetation distributions or greenhouse gases other than CO$_2$, e.g. CH$_4$ or N$_2$O need to be accounted for. From the data over the last 800 kyr, the closest approximation of $S^a$ that can be estimated is $S_{[GHG,LI,AE,VG]}$ including the greenhouse gases from ice cores CO$_2$, CH$_4$ and N$_2$O (GHG), land ice (LI), aerosols (AE) and vegetation cover (VG)  \citep{Koehler2010, Rohling_palaeosens2012} (Fig.~\ref{f:sensitivity_data1}b). 

\begin{figure}
\noindent\includegraphics[width=0.6\textwidth]{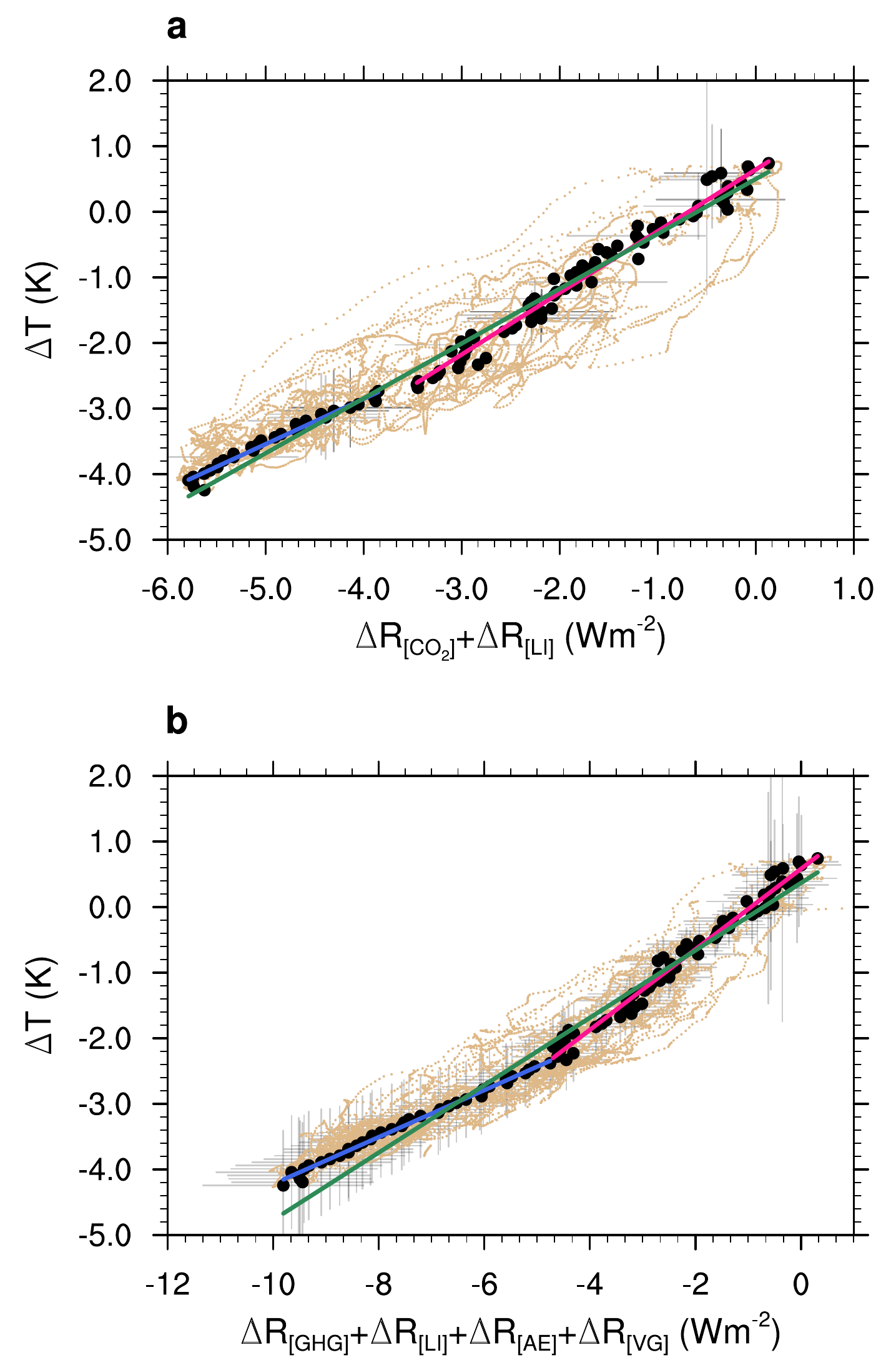}
\caption{Climate sensitivity based on 800 kyr of data. (a) Temperature anomaly considering a global cooling at LGM of $\Delta T=-4.0$~K versus $\Delta R_{[{\rm CO}_2]}+\Delta R_{[LI]}$; (b) Temperature anomaly versus $\Delta R_{[GHG]}+\Delta R_{[LI]}+\Delta R_{[AE]}+\Delta R_{[VG]}$;  Light dots indicate all data points, black thick dots represent the data set divided into 100 temperature bins (bin size $\simeq$~0.05~K) with horizontal and vertical lines denoting the uncertainty limits ($\pm 1\sigma$) for each point; Linear regressions on the binned data using all points (green), only warm data (red) and only cold data (blue), respectively, are calculated. The breakpoint between cold and warm data is determined in SI, section S4.}
\label{f:sensitivity_data2}
\end{figure}

In Fig.~\ref{f:sensitivity_data2} the temperature anomalies are shown versus the radiative perturbations. The uncertainties in both temperature and radiative perturbations as well as the spread in observed values are generally large. In order to estimate a local slope between temperature and radiative forcing, we divide the data into temperature bins and show the average radiative perturbation for each bin together with its uncertainty (black dots in Fig.~\ref{f:sensitivity_data2}). The (local) slopes are then determined using linear regression that 
accounts for errors in both the predictand ($\Delta R_{[X]}$) and the dependent variable ($\Delta T$)  \citep{NumRec}. The regression parameters (y-axis intercept and slope) are returned together with their uncertainties based on the uncertainties of the original data. For each linear regression, we determine the coefficient of determination $r^2$ to assess the explained variance of the fit (SI section S4).

Initially, we perform the regression analysis on the complete binned data set, i.e. we assume no state dependency. Alternatively, the data set is divided into two parts, and for each part the same linear regression analysis is applied. The sum of the two squared residuals is minimised in order to find the optimal breakpoint  \citep{Easterling1995}. 
At this breakpoint we test the significance of the two-phase fit by using a likelihood statistic as in  \citet{Easterling1995} based on the squared residuals of one fit to the whole data set and the squared residuals of the two separate fits. In all cases a F-test reveals that breaking up the data set into two parts gives a statistically significant better fit to the data than only one regression line. Finally, the slopes of the two individual regression lines are in all cases significantly different from each other and from the one-fit regression slope using a student t-test at the 95\% significance level. The result of the linear regression analysis is that taking into account state dependency by dividing the data set in two parts yields generally lower values for the climate sensitivitiy during cold (glacial) periods than during warm intervals (Table~\ref{t:datasens}). 

In the conceptual model of section 2, the dominating fast feedback is the sea ice-albedo feedback, which tends to be stronger during cold periods and therefore leads to higher climate sensitivity during glacial periods. Previous model studies have suggested  that not only the sea-ice albedo feedback \citep{Ritz2011}, but also the short-wave cloud feedbacks \citep{Crucifix2006,Hargreaves2007} or the water vapor and lapse-rate feedbacks \citep{Yoshimori2011} are temperature dependent. Our results here suggest that these other fast feedbacks, which promote a higher sensitivity in warm climates, are stronger than the sea-ice albedo feedback. The sea ice albedo feedback is less effective in warm climates with little or no ice and the water vapour feedback is stronger when there is more moisture in the atmosphere, i.e. in a warm climate with enhanced hydrological cycle. 
To disentangle the contributions of individual feedbacks from the observations, accurate reconstructions of sea-ice, aerosols, clouds etc. are needed. Alternatively, climate models run for at least a few glacial-interglacial cycles could  be used to estimate these contributions. 

\section{Discussion and Conclusions}
%{Correcting for the state dependency of fast feedbacks}

Using the local slope of the relation between temperature anomalies and radiative forcing and assuming that land-ice provides the dominant slow feedback, we estimate the specific sensitivity $S_{[{\rm CO}_2,LI]} = 0.95\pm0.09$~\KWmm\ for global mean temperature anomalies between -2.7 and +0.8~K. 
This value is higher than a previous estimate $S_{[{\rm CO}_2,LI]} = 0.74 \pm 0.28$~\KWmm\ (scaled to the LGM temperature reconstruction as used here) based on the same radiative forcing data \citep{Rohling_palaeosens2012} but neglecting the state dependency of the fast  feedbacks. 

Considering all available forcings the Charney climate sensitivity $S^a$ should be approximated by $S_{[GHG,LI,AE,VG]}$, which we estimate to be $0.61\pm 0.06$~\KWmm\  for warm climates.
Our estimate of $S^a$, however, strongly depends on the scaling of the temperature record to match the most recent reconstruction of LGM cooling \citep{Annan2013}, which might underestimate tropical temperature change \citep{Schmidt2014}. An earlier estimate by \citet{SchneidervonDeimling2006} suggests a LGM cooling of $-5.8\pm 1.4$~K, which leads in our analysis to $S^a = 0.87 \pm 0.08$~\KWmm\ (Table~\ref{t:datasens}, SI section S5). Another estimate based on one climate model and proxy data suggests even less LGM cooling than in \citet{Annan2013} of only $-3.0$ (90\% probability range [-1.7, 3.7]~K, \citep{Schmittner2011}) leading in our analysis to even lower values of $S^a$.

Several factors in our analysis, which at this stage cannot be explicitly taken into account, might influence the estimate for $S^a$: 
(i) In order to match the most recent reconstruction of global mean cooling at the LGM \citep{Annan2013}, we assumed time-independent polar amplification factors for the southern and northern hemisphere \citep{Singarayer2010,MassonDelmotte2006}. Although their uncertainty estimates partly account for a possible time dependency, more information on the relation between high-latitude and global mean temperatures is necessary.
(ii) The efficacy of climate forcings due to varying spatial distribution of radiative forcings can vary over time, which makes it difficult to directly compare the future double $\rm{CO}_2$ experiments with glacial climate forcing \citep{Hansen2005}. 
(iii) Orbital forcing varies over time, and while the varying insolation has been included in the analysis, we did not take into account the dependency of the fast feedbacks on the solar insolation. However, in a previous study \citep{Koehler2010} it was shown for the sea-ice albedo feedback, that the impact of sea-ice area changes between glacial and interglacial states is much larger than the effect of local insolation changes. 
(iv) The equilibrium concept of climate sensitivity might not be adequate in the presence of climate variations on millennial time scales such as the large and rapid changes during Dansgaard-Oeschger events which are believed to be caused by nonlinear processes in the climate system  \citep{Schulz2002,Ganopolski2002, Ditlevsen2009}. For our dataset the equilibrium assumption has been tested \citep{Rohling_palaeosens2012,Koehler2010} and excluding data points from quickly varying periods did not strongly affect the results.

There are three ways to further improve the estimate of $S^a$ from palaeoclimate data:    
(i) Extend the analysis on how climate sensitivity depends on temperature to a wider range  by including past reconstructions of warmer climates (e.g. Pliocene). Indeed, a limitation of the analysis is that our results are based on data of mostly colder than present climate.
(ii) Add more information on the different feedback processes and their dependency on the global mean temperature contributing to the  combined feedback parameter by improved reconstructions. 
(iii) Improve the temperature reconstructions. For the
late Pleistocene including the LGM the uncertainties and differences in $\Delta T$ are still large, in particular for the land-ocean temperature differences and meridional temperature gradients.

In summary, we have provided a novel method to estimate the equilibrium climate sensitivity $S^a$ from a palaeo-data set explicitly accounting for a 
possible state dependency of the fast feedbacks. From 
data (and model-based interpretation) covering the last 800 kyr, we estimate $S^a = 0.61\pm0.06$~\KWmm (given the latest LGM cooling reconstruction of $-4.0$~K) valid for global mean temperatures between 2.3~K colder to 0.8~K warmer than the preindustrial climate. Due to the large uncertainty of LGM temperature reconstructions this value may be higher up to $S^a_{\Delta T_{LGM}=-5.8K} = 0.87\pm0.08$~\KWmm.
This corresponds to an equilibrium global mean surface warming of $2.0-3.5$~K  for $2\times {\rm CO}_2$. 
These estimates can be further improved if more accurate temperature reconstructions and better estimates 
of radiative forcing due to slow feedbacks become available.   

%%% End of body of article:

%%%%%%%%%%%%%%%%%%%%%%%%%%%%%%%%
%% Optional Appendix goes here
%
 % \appendix 
 % resets counters and redefines section heads
% but doesn't print anything.
% After typing \appendix
%
% will show
% Appendix A: Here Is Appendix Title
%
%%%%%%%%%%%%%%%%%%%%%%%%%%%%%%%%%%%%%%%%%%%%%%%%%%%%%%%%%%%%%%%%
%
% Optional Glossary or Notation section, goes here
%
%%%%%%%%%%%%%%
% Glossary is only allowed in Reviews of Geophysics
% \section*{Glossary}
% \paragraph{Term}
% Term Definition here
%
%%%%%%%%%%%%%%
% Notation -- End each entry with a period.
% \begin{notation}
% Term & definition.\\
% Second term & second definition.\\
% \end{notation}
%%%%%%%%%%%%%%%%%%%%%%%%%%%%%%%%%%%%%%%%%%%%%%%%%%%%%%%%%%%%%%%%
%
%  ACKNOWLEDGMENTS

\begin{acknowledgments}
The analysis of the data-based changes in climate sensitivity for the last 800~kyr used the following public available data sets from the NOAA National Climatic Data Center (http://www.ncdc.noaa.gov/paleo/paleo.html): $\Delta T_{NH}$ \citep{Bintanja2005}, $\Delta T_{ANT}$ \citep{Jouzel2007}, ${\rm CO}_2$ \citep{Petit1999,Monnin2001,Siegenthaler2005,Luethi2008}, sea level reconstructions \citep{Bintanja2005}. Further information on land-ice distribution \citep{Bintanja2005} and on calculated radiative forcing to obtain $\Delta R_{[LI]}$ \citep{Koehler2010} were necessary for the data-based analysis plotted in Figs. 2 and 3. 
\end{acknowledgments}

\end{article}

%
%
%% Enter Figures and Tables here:
%
% DO NOT USE \psfrag or \subfigure commands.
%
% Figure captions go below the figure.
% Table titles go above tables; all other caption information
%  should be placed in footnotes below the table.
%
%----------------
% EXAMPLE FIGURE
%
% \begin{figure}
% \noindent\includegraphics[width=20pc]{samplefigure.eps}
% \caption{Caption text here}
% \label{figure_label}
% \end{figure}

\clearpage

%
% ---------------
% EXAMPLE TABLE
%
%\begin{table}
%\caption{Time of the Transition Between Phase 1 and Phase 2\tablenotemark{a}}
%\centering
%\begin{tabular}{l c}
%\hline
% Run  & Time (min)  \\
%\hline
%  $l1$  & 260   \\
%  $l2$  & 300   \\
%  $l3$  & 340   \\
%  $h1$  & 270   \\
%  $h2$  & 250   \\
%  $h3$  & 380   \\
%  $r1$  & 370   \\
%  $r2$  & 390   \\
%\hline
%\end{tabular}
%\tablenotetext{a}{Footnote text here.}
%\end{table}

% See below for how to make sideways figures or tables.

\clearpage

\begin{sidewaystable}
\caption{Overview of $S_{\rm{[CO_2},LI]}$ and $S_{[GHG,LI,AE,VG]}$ (K (W m$^{-2}$)$^{-1}$) calculated from data of the last 800 kyr depending on whether or not the background climate state of the fast feedbacks are considered and on how temperature time series are normalized to fit reconstructed global cooling at LGM ($\Delta T_{\rm LGM}$). We use the most recent extensive reconstruction of the LGM climate by \citet{Annan2013} as standard. The spread in values is assessed by using one other reconstruction indicating a considerably stronger LGM cooling of $-5.8$~K \citep{SchneidervonDeimling2006}.}
\centering
\begin{tabular}{|c|c|c|c||c|c|c|}
\hline
\multicolumn{7}{|c|}{$\Delta T_{\rm LGM}=-4.0 \pm 0.8$~K \citep{Annan2013}}\\
\hline
                       &  \multicolumn{3}{c||}{$S_{\mathrm{[CO_2},LI]}$ } & \multicolumn{3}{c|}{$S_{\mathrm{[GHG,LI,AE,VG]}}$ }\\
\hline
$\Delta$T range & all & {\color{blue} cold} & {\color{red} warm}  & all & {\color{blue} cold} & {\color{red} warm}\\
considered (K) & -4.3 -- 0.8 & {\color{blue} -4.3 -- -2.7} & {\color{red} -2.7 -- 0.8} & -4.3 -- 0.8 & {\color{blue} -4.3 -- -2.3} & {\color{red} -2.3 -- 0.8}\\
\hline
%S$\pm\sigma$(S) & 0.84$\pm$0.05\tablefootnote{Fig.~3a green line.} & {\color{blue} 0.69$\pm$0.24\tablefootnote{Fig.~3a blue line.}} & {\color{red} 0.94$\pm$0.09\tablefootnote{Fig.~3a red line.}} &  0.51$\pm$0.03\tablefootnote{Fig.~3b green line.} & {\color{blue} 0.36$\pm$0.08\tablefootnote{Fig.~3b blue line.}} & {\color{red} 0.61$\pm$0.06\tablefootnote{Fig.~3b red line.}}\\
S$\pm\sigma$(S) & 0.84$\pm$0.05\tablefootnote{Fig.~\ref{f:sensitivity_data2}a green line.} & {\color{blue} 0.69$\pm$0.24\tablefootnote{Fig.~\ref{f:sensitivity_data2}a blue line.}} & {\color{red} 0.94$\pm$0.09\tablefootnote{Fig.~\ref{f:sensitivity_data2}a red line.}} &  0.51$\pm$0.03\tablefootnote{Fig.~\ref{f:sensitivity_data2}b green line.} & {\color{blue} 0.36$\pm$0.08\tablefootnote{Fig.~\ref{f:sensitivity_data2}b blue line.}} & {\color{red} 0.61$\pm$0.06\tablefootnote{Fig.~\ref{f:sensitivity_data2}b red line.}}\\
(\KWmm) & & & & & & \\
\hline
previous study & 0.74$\pm$0.28 & & &  0.47$\pm$0.17 & & \\
\hline
\hline
\multicolumn{7}{|c|}{$\Delta T_{\rm LGM}=-4.0 ... -5.8$~K \citep{Annan2013,SchneidervonDeimling2006}}\\
\hline
                       &  \multicolumn{3}{c||}{$S_{\mathrm{[CO_2},LI]}$ } & \multicolumn{3}{c|}{$S_{\mathrm{[GHG,LI,AE,VG]}}$ }\\
\hline
$\Delta$T range & all & {\color{blue} cold} & {\color{red} warm}  & all & {\color{blue} cold} & {\color{red} warm}\\
considered     &  &  &  & &  & \\
\hline
S range & 0.79 -- 1.23 & {\color{blue} 0.45 -- 1.25} & {\color{red} 0.85 -- 1.44} & 0.48 -- 0.76 & {\color{blue} 0.28 -- 0.60 } & {\color{red} 0.55 -- 0.95}\\
(\KWmm) & & & & & & \\
\hline

\end{tabular}
\label{t:datasens}
\end{sidewaystable}

%TC:endignore

\end{document}